# EVALUASI PENERAPAN MODEL PEMBELAJARAN INKUIRI TERBIMBING DALAM PEMBELAJARAN KIMIA : SUATU TINJAUAN SISTEMATIS LITERATUR

Ainayya Almira[1], Anisah Rachmawati[2], Insi Norma Jelita[3], Yoan Nurlaili[4].

Pendidikan Kimia, Universitas Negeri Malang, Kota Malang, 65114.

Email: ainayya.almira.2103316@students.um.ac.id,

anisah.rachmawati.2103316@students.um.ac.id,

insi.norma.2103316@students.um.ac.id, yoan.nurlaili.2103316@students.um.ac.id

**Abstrak**

Tujuan penelitian ini memberikan wawasan kepada guru dan peneliti pendidikan kimia mengenai efektivitas model pembelajaran inkuiri terbimbing serta memberikan arahan untuk penelitian lebih lanjut dalam bidang ini. Metode penelitian yang digunakan dalam artikel ini adalah *Systematic Literature Review (SLR)*, Untuk membantu menyusun dan mengevaluasi berbagai penelitian terkait model pembelajaran inkuiri terbimbing. Instrumen yang digunakan dalam penelitian ini adalah menyajikan hasil tinjauan pustaka dari berbagai artikel yang membahas penerapan model ini dalam pembelajaran kimia dengan menggali definisi, penerapan, kelebihan, kelemahan, dan efektivitas model pembelajaran inkuiri terbimbing dalam pembelajaran kimia. Hasil penelitian menunjukkan bahwa penerapan model ini dapat dilakukan baik dalam aspek teoritis maupun praktek pembelajaran kimia. Kelebihan model inkuiri terbimbing melibatkan siswa secara aktif, meningkatkan kemandirian belajar, dan memberikan kesempatan siswa untuk berdiskusi dan menemukan jawaban sendiri. Siswa yang belajar dengan model ini cenderung memiliki prestasi belajar yang lebih tinggi. Namun, terdapat juga kelemahan, seperti waktu yang dibutuhkan untuk mengimplementasikan model ini dan kendala dalam menangani siswa yang belum terbiasa dengan pendekatan ini.

**Kata kunci**: Inkuiri terbimbing; *Systematic Literature Review (SLR)*; Efektivitas inkuiri terbimbing

*Abstract*

*The aim of this research is to provide insight to chemistry education teachers and researchers regarding the effectiveness of the guided inquiry learning model and provide direction for further research in this field. The research method used in this article is Systematic Literature Review (SLR), to help compile and evaluate various research related to the guided inquiry learning model. The instrument used in this research is to present the results of a literature review of various articles discussing the application of this model in chemistry learning by exploring the definition, application, strengths, weaknesses and effectiveness of the guided inquiry learning model in chemistry learning. The research results show that the application of this model can be carried out both in the theoretical and practical aspects of chemistry learning. The advantages of the guided inquiry model involve students actively, increase learning independence, and provide students with the opportunity to discuss and find their own answers. Students who study with this model tend to have higher learning achievements. However, there are also disadvantages, such as the time required to implement this model and obstacles in dealing with students who are not yet familiar with this approach.*

***Keyword:*** *guided inquiry, Systematic Literature Review (SLR), effectiveness of the guided inquiry*

**Pendahuluan**

Kimia merupakan suatu pelajaran yang mempelajari terkait susunan, sifat dan reaksi dari suatu unsur, serta perubahan suatu materi dan energi yang menyertainya. Kimia sangat erat hubungannya dengan kehidupan sehari-hari. Oleh karena itu mata pelajaran kimia menjadi sangat penting peranannya di dalam dunia pendidikan. Banyak peristiwa di alam yang sering ditemui dapat dipelajari di dalam ilmu kimia (Viandhika, 2015). Kimia menjadi salah satu mata pelajaran yang dianggap sulit oleh sebagian siswa sekolah menengah atas (SMA). Hal ini dikarenakan konsep kimia memiliki sifat abstrak dan kompleks, sehingga memerlukan pemahaman yang mendalam terhadap konsep dasar. Keabstrakan konsep kimia tersebut berpotensi menyebabkan kesulitan bagi siswa untuk mempelajarinya. Kesulitan belajar kimia dapat disebabkan oleh kurangnya pemahaman siswa atas konsep dasar kimia yang bersifat abstrak atau tidak dapat dilihat dan dipahami dengan panca indera. Materi di dalam ilmu kimia yang mencakup perhitungan dan hafalan menjadikan banyak dari mereka yang merasa kesulitan dalam memahami dan mengaplikasikan rumus-rumus selama pembelajaran kimia berlangsung (Ratna, 2014). Kutipan harus ditulis dengan menggunakan format bodynote seperti (Uwuigbe & Ajibolade, 2013), (Wang, 2016), (Muttakin et al., 2015) dan relevan dengan daftar Pustaka/ Bibliografi (disarankan menggunakan Aplikasi Mendeley).

Keabstrakan konsep kimia dapat dipahami dengan mudah oleh siswa dengan melibatkan berbagai representasi dalam proses pembelajaran kimia (Wiyarsi, 2018). Representasi tersebut dapat terlaksana melalui berbagai model pembelajaran yang sesuai dengan materi yang akan disampaikan. Model pembelajaran yang tepat tentunya dapat menjembatani pemahaman siswa terhadap keabstrakan konsep kimia. Salah satu model pembelajaran yang sering digunakan dalam pembelajaran kimia yakni inkuiri terbimbing. Model pembelajaran inkuiri terbimbing adalah model yang memungkinkan peserta didik untuk bergerak dalam mengidentifikasi masalah, merumuskan masalah, hipotesis, pengumpulan data, verifikasi hasil, dan penarikan kesimpulan (Matthew dan Igharo, 2013).

Pola pembelajaran inkuiri terbimbing secara langsung melibatkan siswa dalam proses pembelajaran. Sehingga akan mendorong siswa berperan lebih aktif. Metode Inkuiri ini ditentukan oleh keseluruhan aspek pengajaran di kelas, proses keterbukaan dan peran aktif siswa. Pada prinsipnya keseluruhan proses pembelajaran akan membantu siswa menjadi percaya diri dan yakin pada kemampuan intelektualnya sendiri untuk terlibat secara aktif. Dalam pembelajaran inkuiri terbimbing siswa dapat memperoleh konsep materi dengan mengkonstruksi konsep dengan mandiri (Putri, 2020). Dengan demikian penguasaan konsep kimia siswa yang dibelajarkan dengan model pembelajaran inkuiri terbimbing lebih tinggi daripada siswa yang dibelajarkan dengan pembelajaran konvensional.

Artikel tinjauan pustaka ini menyajikan hasil kajian dari beberapa artikel yang berfokus pada pembelajaran kimia menggunakan inkuiri terbimbing. Laporan-laporan yang dikaji dalam tinjauan pustaka ini diperoleh melalui sintesis sistematis artikel-artikel pada tahun 2014 hingga tahun 2022.
Melalui artikel tinjauan pustaka ini diharapkan para guru dan peneliti pendidikan kimia mendapatkan wawasan dan informasi tentang model pembelajaran inkuiri terbimbing, pengaruh penerapannya terhadap pemahaman konsep dan hasil belajar siswa dalam pembelajaran kimia, serta bagaimana kesesuaiannya dalam pembelajaran kimia secara teori maupun praktikum. Selain itu juga memberikan arahan bagi peneliti pendidikan kimia untuk penyelidikan lebih lanjut. Pertanyaan penelitian yang memandu penulisan artikel ini adalah:

1. Apa yang dimaksud dengan model pembelajaran inkuiri terbimbing?
2. Bagaimana penerapan model pembelajaran inkuiri terbimbing dalam proses pembelajaran kimia secara teori maupun praktikum?
3. Bagaimana pengaruh penerapan model pembelajaran inkuiri terbimbing terhadap pemahaman konsep dan hasil belajar siswa dalam pembelajaran siswa?

**Metode Penelitian**

Metode yang digunakan dalam artikel ini adalah *Systematic Literature Review* (SLR). Systematic Literature Review merupakan istilah yang digunakan untuk merujuk pada metodologi penelitian atau riset tertentu dan pengembangan yang dilakukan untuk mengumpulkan serta mengevaluasi penelitian yang terkait pada fokus topik tertentu (Lusiana, 2014). Dengan penggunaan Metode SLR dapat dilakukan review dan identifikasi jurnal secara sistematis yang pada setiap prosesnya mengikuti langkah-langkah atau protokol yang telah ditetapkan (B. Kitchenham, 2009). Pengambilan artikel dan analisis dilakukan pada tanggal 16 November sampai dengan 7 Desember 2023. Dalam penelitian ini kata kunci yang digunakan adalah "Model pembelajaran inkuiri", "Model pembelajaran inkuiri pada pelajaran kimia", "Model pembelajaran kimia berbasis inkuiri terbimbing", "Penerapan model pembelajaran inkuiri terbimbing", dan "Pengaruh model pembelajaran inkuiri terbimbing".

Setelah mencari kata kunci, peneliti melanjutkan memilah artikel seperti pada **Gambar 1** yang sesuai dengan kriteria inklusi sebagai berikut: (1) artikel berasal dari jurnal pendidikan; (2) berkaitan dengan pembelajaran menggunakan inkuiri dan dampak pembelajaran menggunakan inkuiri; (3) tahun penerbitan artikel dari tahun 2010 sampai dengan tahun 2023. Berdasarkan hasil pencarian, terdapat 40 artikel yang sesuai dengan kriteria inklusi diatas. Namun setelah dikerucutkan lagi berdasarkan judul yang berkaitan dengan inkuiri terbimbing, didapatkan hanya 26 artikel. Sedangkan 14 artikel lainnya mengangkat judung inkuiri saja. Kemudian dilakukan analisa pada bagian abstrak dan isi, diperoleh 13 artikel dengan teknik pengambilan data dan inti dari pembahasan terkait pengaruh model pembelajaran yang sama. Dari 13 artikel inilah akan didapatkan informasi yang sesuai dengan topik pembahasan peneliti.

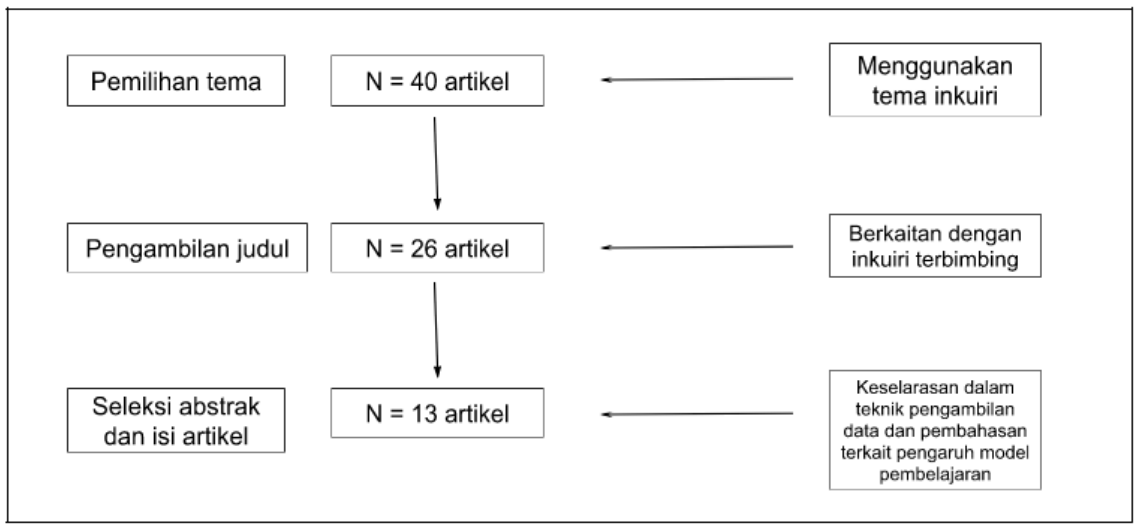

**Gambar 1 : proses seleksi artikel**

**Hasil dan Pembahasan**

Hasil penelitian yang didapatkan pada seleksi bertahap, didapatkan 13 artikel mengenai model pembelajaran inquiry terbimbing dalam pembelajaran kimia yang dipublikasikan pada tahun yang berbeda - beda, pada rentang tahun 2014 sampai dengan 2022 seperti pada **Gambar 2**. Artikel-artikel tersebut sebagian besar membahas mengenai keefektivitasan pembelajaran inquiry terbimbing dalam pembelajaran kimia.

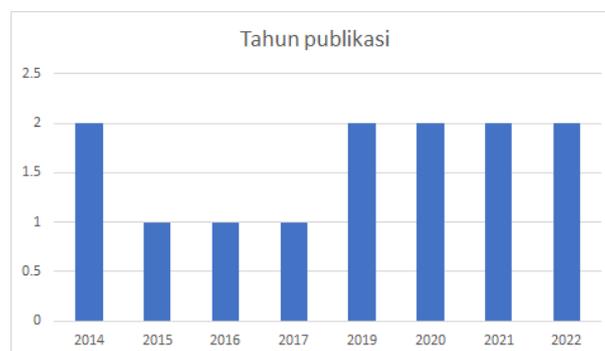

**Gambar 2 : Tahun publikasi artikel model inquiry terbimbing**

Pengkajian pada 13 artikel ini mengerucut dari total keseluruhan 40 artikel, diperoleh analisis mengenai model pembelajaran inkuiri terbimbing, penerapan model pembelajaran inkuiri terbimbing dalam proses pembelajaran kimia, pengaruh model pembelajaran inkuiri terbimbing terhadap pemahaman konsep dan hasil belajar siswa. Kajian pertama dari artikel terdapat pada **Tabel 1** mengenai definisi model pembelajaran inkuiri terbimbing. Model pembelajaran inkuiri terbimbing mencangkup proses berpikir kritis dan analisis siswa dalam memecahkan suatu masalah.

**Tabel 1. Model pembelajaran inkuiri terbimbing**

| Definisi | Penulis |
| --- | --- |
| Model pembelajaran inkuiri terbimbing berpusat pada siswa dan proses berpikir kritis dan analitis siswa untuk memecahkan suatu masalah. | (Eralita & Sulistyo Saputro, 2015) ; (Kelarutan dkk., 2017); (Hariyadi & Rahayu, 2016) |
| Model pembelajaran inkuiri terbimbing melibatkan secara aktif dalam proses pengembangan pengetahuan, dimana siswa mengidentifikasi masalah sampai penarikan kesimpulan, | (Pratiwi Pane dkk., 2021); (Sundami & Azhar, 2019); (Novitasari, 2020); (Husain dkk., 2022); (Wardani & Firdaus, 2019) |

Hasil penelitian dalam review didapatkan bahwa model pembelajaran inkuiri terbimbing memiliki kata kunci berupa siswa aktif, berpikir kritis, dan identifikasi masalah. Dari setiap definisi yang tertera pada 13 artikel mengandung maksud yang sama. Selanjutnya mengkaji mengenai penerapan model pembelajaran inkuiri terbimbing dalam proses pembelajaran kimia. Pada **Tabel 2** didapatkan analisis bahwa penerapan model pembelajaran inkuiri terbimbing dapat diterapkan di pembelajaran kimia dalam aspek teoritis maupun prakteknya. Artikel yang direview menggunakan materi yang dapat dilakukan dengan kedua pendekatan yaitu pendekatan teoritis dan praktek seperti hidrokarbon, kesetimbangan kimia, ikatan kimia, larutan penyangga dan hasil kali kelarutan.

**Tabel 2. Penerapan model pembelajaran inkuiri terbimbing**

| Penerapan | Sinta | Penulis |
|---|---|---|
| Teoritis | Model pembelajaran inkuiri terbimbing terdiri dari beberapa langkah pembelajaran diantaranya, orientasi pada masalah, merumuskan masalah, mengajukan hipotesis, mengumpulkan data, menguji hipotesis dan merumuskan kesimpulan. | (Husain dkk., 2022); (Novitasari, 2020); (Qurrotu & Ismono, 2020); (Sundami & Azhar, 2019); (Wardani & Firdaus, 2019); (Muhali dkk., 2021); (Eralita & Sulistyo Saputro, 2015); (Firdausi, 2014); (Budi Santoso & Rante Suparman, 2017); (Noer Syamsiyah, 2022); (Pratiwi Pane dkk., 2021). |
| Praktek | Sintak meliputi menyajikan pertanyaan dan masalah, membuat hipotesis, merancang percobaan, menguji hipotesis melalui percobaan untuk memperoleh informasi, mengumpulkan serta analisis data, hingga membuat kesimpulan (Trianto, 2009) | (Pratiwi Pane dkk., 2021) ; (Muhali dkk., 2021) ; (Fakayode, 2014) ; |

Berdasarkan analisis artikel, sintak yang digunakan dalam aspek teoritis dan praktik memiliki kesamaan yaitu siswa aktif dalam pemecahan masalah hingga menarik kesimpulan hasil analisisnya. Dengan begitu analisis dilanjutkan dengan kelebihan model pembelajaran inkuiri terbimbing terhadap kedua aspek pendekatan dalam pembelajaran kimia yaitu aspek teoritis dan praktek. Berdasarkan **Tabel 3**, penerapan model inkuiri terbimbing pada pembelajaran kimia aspek teoritis memberikan segudang manfaat terutama dalam keaktifan dan pemahaman konsep bagi peserta didik.

**Tabel 3. Kelebihan penerapan model inkuiri terbimbing dalam aspek teoritis**

| Kelebihan | Penulis |
|---|---|
| 1. Dapat memberikan kesempatan dan dorongan alami pada peserta didik untuk melakukan eksplorasi.<br>2. Peningkatan kemandirian belajar dari peserta didik.<br>3. Memberikan kesempatan agar dapat berdiskusi mencari dan menemukan jawaban sendiri dari rumusan masalah yang telah dibuat, sehingga diharapkan peserta didik dapat menumbuhkan sikap percaya diri dan dapat berperan aktif.<br>4. Memberi kesempatan lebih banyak untuk merefleksikan pembelajaran, mendapat pemahaman yang lebih dalam, sehingga mampu mengoptimalkan kemampuan kognitif dan psikomotor. | (Wardani & Firdaus, 2019) |
| Model pembelajaran inkuiri terbimbing memiliki keuntungan bagi siswa yaitu mengembangkan kemampuan sosial, membaca dan keterampilan berbahasa siswa, membangun pemahaman mereka sendiri mengenai hal mereka selidiki, siswa bebas dalam meneliti dan belajar, serta memiliki keterlibatan secara langsung. | (Noer Syamsiyah, 2022) |
| Model pembelajaran yang diterapkan dalam penelitian ini telah memiliki sintaks yang mampu meningkatkan HOTS siswa. | (Nur Indah Firdausi, 2014) |
| 1. Peserta didik kelas eksperimen melakukan praktikum dengan rasa ingin tahu yang besar pada sebagian besar peserta didik di kelas tersebut. Sedangkan pada kelas kontrol hanya sebagian kecil peserta didik yang aktif dan melakukan praktikum.<br>2. Peserta didik dapat lebih lama mengingat informasi pengetahuan yang ditemukan.<br>3. Konsep pada materi pelajaran hukum dasar kimia dapat terbentuk dengan baik | (Sumarni S., Bimo Budi Santoso., Achmad Rante Suparman. 2017) |

| Kelebihan | Penulis |
|---|---|
| 4. Keterampilan proses sains akan meningkat sehingga pembelajaran di dalam kelas lebih aktif.<br>5. Melalui pembelajaran dengan penemuan konsep maka siswa lebih dapat memahami materi.<br>6. Dapat meningkatkan prestasi belajar. | |
| Penggunaan pembelajaran Inkuiri Terbimbing berpengaruh signifikan terhadap prestasi belajar, kreativitas, dan motivasi belajar kognitif, afektif, dan psikomotorik siswa | (Qurrotu & Ismono, 2020) |

Setelah pemaparan kelebihan dari aspek teoritis, dilanjutkan dengan kelebihan dari aspek prakteknya. Berdasarkan artikel yang direview, teknik GILEs sangat membantu siswa dalam memahami materi yang akan dipraktikumkan. Dengan teknik tersebut dapat menunjang model pembelajaran inkuiri terbimbing. Pernyataan tersebut disajikan dalam **Tabel 4** kelebihan penerapan model inkuiri terbimbing, ditinjau dari aspek prakteknya.

**Tabel 4. Kelebihan penerapan model inkuiri terbimbing dalam aspek praktek**

| Kelebihan | Penulis |
|---|---|
| 1. Efektif dalam membuat beberapa siswa tertarik dan siswa lainnya lebih menerima pendekatan GILE<br>2. Meningkatkan pembelajaran siswa dan kemampuan berpikir kritis | (Fakayode, 2014) |
| Pemahaman konsep<br>1. Memberikan kesempatan pada siswa untuk dapat melakukan konstruksi konsepnya sendiri, sehingga pemahaman akan konsep materi yang dibelajarkan bertahan lebih lama.<br>2. Menuntut siswa dapat melakukan kegiatan berpikir seperti perumusan masalah dan pengujian hipotesis yang dikemukakan, yang berdampak pada aktifnya siswa dalam membangun konsepnya sendiri. | (Muhali dkk., 2021) |

| | Keterampilan metakognitif<br>1. Memberikan kesempatan kepada siswa dalam membuat perencanaan dengan lebih baik, dan memfokuskan siswa terhadap yang dipelajari.<br>2. Memberikan kesempatan mendapatkan informasi dengan lebih efisien, juga menyadarkan siswa akan kontrol berpikirnya, sehingga tahu apa yang dilakukan dan bagaimana melakukannya.<br>3. membuat siswa lebih sadar akan proses berpikir yang dilakukannya.<br>4. pembelajaran yang dilakukan di kelas eksperimen menekankan pada pembelajaran aktif dan pengembangan keterampilan berpikir. | |
|---|---|---|

Dari banyaknya kelebihan yang sudah dipaparkan terdapat pula kelemahan dari aspek teoritis maupun aspek praktek. Dalam artikel yang direview ada banyak kelemahan seperti pada **Tabel 5** yang menjadikan hal itu sebagai masalah baru dalam aspek teoritis maupun praktek dalam model pembelajaran inkuiri terbimbing.

**Tabel 5. Kelemahan penerapan model inkuiri terbimbing dalam aspek teoritis dan praktek**

| Aspek | Kelemahan | Penulis |
|---|---|---|
| Teoritis | 1. Peserta didik belum terbiasa menggunakan model inkuiri terbimbing menyelesaikan permasalahan sehingga guru agak sulit dalam menerapkan model ini<br>2. Dalam mengimplementasikan model inkuiri terbimbing memerlukan waktu yang panjang sehingga guru sulit menyesuaikan dengan waktu yang telah ditentukan. | (Wardani & Firdaus, 2019); (Noer Syamsiyah, 2022); (Firdausi, 2014); (Budi Santoso & Rante Suparman, 2017); (Hariyadi & Rahayu, 2016) |

|  | 3. Pada pelaksanaannya perlu 82memperhatikan karakteristik siswa, ketersediaan sumber belajar, dan alokasi waktu yang tersedia.<br>4. Struktur tujuan kooperatif yang menciptakan suatu situasi dimana satu-satunya cara agar anggota kelompok dapat mencapai tujuan pribadi mereka hanya apabila kelompoknya berhasil.<br>5. Siswa harus mencari sendiri pelajaran atau materi pembelajaran<br>6. Masih mengalami kesulitan dalam menyelesaikan pertanyaan-pertanyaan dengan tingkat kesulitan C4, C5, dan C6.<br>7. Suasana kelas menjadi ramai terutama saat proses diskusi dan dapat mengganggu proses belajar dan konsentrasi peserta didik<br>Menganggap bahwa mata pelajaran kimia adalah mata pelajaran yang sulit dan susah untuk dipahami peserta didik. |  |
|---|---|---|
| Praktek | 1. Memerlukan lebih banyak persediaan dan sumber daya dibandingkan eksperimen tradisional.<br>Memakan banyak waktu. | (Fakayode, 2014) |

Selanjutnya pada **Tabel 6** yaitu kajian terakhir mengenai efektivitas pembelajaran inkuiri terbimbing terhadap aspek teoritis dan aspek praktek. Efektivitas pembelajaran ini berkaitan dengan pemahaman konsep dan hasil belajar siswa. Dari beberapa artikel yang telah ditinjau, didapatkan bahwa penerapan model inkuiri terbimbing sangat efektif terbukti dari beberapa artikel yang menyebutkan efektivitas dari model pembelajaran tersebut.

**Tabel 6. Efektivitas penerapan model inkuiri terbimbing dalam aspek teoritis dan praktek**

| Aspek | Efektivitas | Penulis |
|---|---|---|
| Teoritis | Pembelajaran inkuiri terbimbing dan learning cycle 5E merupakan pembelajaran dengan konsep konstruktivisme, yang membawa siswa untuk membangun dan mengeksplorasi semua kemampuan berpikir dan keterampilan yang dimiliki oleh siswa. Pembahasan di atas menunjukkan bahwa model inkuiri terbimbing berbasis blended learning mampu meningkatkan kemampuan kognitif pada peserta didik kelas eksperimen daripada penerapan model konvensional berbasis blended learning pada kelas kontrol. Hasil pengamatan pada siklus I dan siklus II mengalami peningkatan hasil belajar dan juga aktivitas baik bagi guru maupun bagi siswa dalam kegiatan belajar mengajar dan mencapai ketuntasan. dapat disimpulkan terdapat peningkatan hasil belajar siswa kelas X materi ikatan kimia melalui penerapan model pembelajaran tipe inkuiri pada SMA Negeri 10 pandeglang. Hal ini disebabkan adanya peningkatan keaktifan dan keterampilan siswa sehingga mengakibatkan efek yang positif terhadap pemahaman siswa dalam mempelajari materi ajar. Hasil belajar kognitif peserta didik yang menerima pembelajaran dengan model Inkuiri Terbimbing lebih tinggi dibandingkan peserta didik yang menerima pelajaran dengan model konvensional. | (Eralita & Sulistyo Saputro, 2015) (Wardani & Firdaus, 2019) (Noer Syamsiyah, 2022) (Budi Santoso & Rante Suparman, 2017) (Hariyadi & Rahayu, 2016) (Husain dkk., 2022) (Sundami & Azhar, 2019) |

|  | Keterampilan proses siswa yang dibelajarkan dengan model pembelajaran inkuiri terbimbing berbasis lingkungan lebih tinggi daripada siswa yang dibelajarkan dengan pembelajaran konvensional. Penguasaan konsep IPA siswa yang dibelajarkan dengan model pembelajaran inkuiri terbimbing berbasis lingkungan lebih tinggi daripada siswa yang dibelajarkan dengan pembelajaran konvensional. peserta didik yang dibelajarkan dengan model Inkuiri Terbimbing memiliki rata-rata nilai yang lebih tinggi dibandingkan dengan model pembelajaran konvensional. pembelajaran dengan model inkuiri terbimbing mampu meningkatkan pemahaman dan penguasaan materi larutan penyangga melalui tahapan inkuiri terbimbing Model pembelajaran inkuiri terstruktur dibandingkan dengan siswa yang menggunakan inkuiri terbimbing pada konsep fotosintesis diperoleh bahwa keterampilan proses sains siswa yang menggunakan inkuiri terstruktur lebih tinggi dibandingkan keterampilan siswa yang menggunakan model pembelajaran inkuiri terbimbing. |  |
|---|---|---|
| Praktek | Persentase hasil ketuntasan belajar melalui model praktikum berbasis proyek dan inquiry mencapai ketuntasan tinggi, yaitu sebesar 93,3% tergolong kategori baik sekali, dan siswa yang tidak tuntas sebesar 6,7%. Hasil ketuntasan belajar menunjukkan bahwa model praktikum berbasis proyek dan inquiry efektif digunakan pada materi asam basa. | (Pratiwi Pane dkk., 2021) |

**Kesimpulan**

Hasil penelitian yang didapatkan, inkuiri terbimbing merupakan model pembelajaran yang melibatkan peserta didik untuk lebih berperan aktif dengan cara berpikir kritis dalam pemecahan masalah. Dengan begitu pengetahuan peserta didik dapat berkembang. Penerapan model pembelajaran inkuiri terbimbing dapat diterapkan di pembelajaran kimia dalam aspek teoritis maupun prakteknya seperti yang tertera pada beberapa artikel yang direview. Meski demikian, masih terdapat beberapa kekurangan yang perlu diperhatikan. Selain itu, model pembelajaran ini juga cukup berpengaruh terhadap pemahaman konsep siswa serta hasil belajar terkait materi kimia yang dibelajarkan. Sehingga dapat disimpulkan bahwa model pembelajaran inkuiri terbimbing cukup efektif diterapkan pada proses pembelajaran kimia secara teoritis maupun praktek.

**Bibliografi**